\documentclass[useAMS,usenatbib,usegraphicx]{mn2e}
\usepackage{psfig}   
\usepackage{graphicx}
\usepackage{amssymb}
\usepackage{subfigure}

\title[On the mass segregation of stars and brown dwarfs in Taurus]{On the mass segregation of stars and brown dwarfs in Taurus}

\author[R.~J.~Parker, J.~Bouvier, S.~P.~Goodwin, E.~Moraux, R.~J.~Allison, S.~Guieu and M.~G{\"u}del]{
  Richard J.~Parker$^{1,2}$\thanks{E-mail: rparker@phys.ethz.ch},
  Jerome Bouvier$^3$, Simon P.~Goodwin$^2$, Estelle Moraux$^3$, \newauthor \hspace*{0.03cm} Richard J.~Allison$^2$, Sylvain Guieu$^4$ and Manuel G{\"u}del$^5$ \vspace*{0.1cm}\\
   $^1$ Institute for Astronomy, ETH Z{\"u}rich, Wolfgang-Pauli-Strasse 27, 8093, Z{\"u}rich, Switzerland \\
   $^2$ Department of Physics and Astronomy, University of Sheffield,
    Sheffield, S3 7RH, UK\\
   $^3$ Laboratoire d'Astrophysique de Grenoble, Observatoire de Grenoble, BP 53, 38041 Grenoble, Cedex 9, France\\
   $^4$ European Southern Observatory, Alonso de Cordova 3107, Vitacura, Santiago, Chile \\ 
   $^5$ Department of Astronomy, University of Vienna, T{\"u}rkenschanzstra{\ss}e 17, Vienna, A-1180, Austria}

\begin{document}

\date{Accepted for publication in MNRAS}
                             
\pagerange{\pageref{firstpage}--\pageref{lastpage}} \pubyear{2010}

\maketitle

\label{firstpage}

\begin{abstract}
We use the new minimum spanning tree (MST) method to look for mass
segregation in the Taurus association.  The method computes the ratio
of MST lengths of any chosen subset of objects, including the most
massive stars  and brown dwarfs, to the MST lengths of
random sets of stars and brown dwarfs in the cluster. This  mass segregation ratio
($\Lambda_{\rm MSR}$) enables a quantitative measure of the spatial distribution of
high-mass and low-mass stars, and brown dwarfs to be made in Taurus.

We find that the most massive stars in Taurus are \emph{inversely mass
  segregated}, with $\Lambda_{\rm MSR} = 0.70 \pm 0.10$  ($\Lambda_{\rm MSR}$ = 1 corresponds to no
mass segregation), which differs from the strong mass segregation
signatures found in  more dense and massive clusters such as
Orion. The brown dwarfs in Taurus are not mass segregated, although we
find evidence that some low-mass stars are, with an $\Lambda_{\rm MSR} = 1.25 \pm 0.15$. Finally, we compare our results to previous  measures of
the spatial distribution of stars and brown dwarfs in Taurus, and briefly 
discuss their implications.
\end{abstract}


\section{Introduction}

The Taurus association is a nearby young cluster \citep[at $140$\,pc,
with an age of $\sim 1$\,Myr;][]{Kenyon94}, still in the process of forming
stars from its natal molecular cloud. It contains relatively few stars
($<$~400), of which most are contained within  several main
aggregates \citep[e.g.][]{Gomez93,Kenyon08}. Star formation in Taurus
appears to be occurring along three  parallel filaments, with the
central filament coincident on the main region of aggregates 
\citep[e.g.][]{Ungerechts87}.

Taurus has a spatial extent of $\sim$ 30\,pc \citep{Palla02}, and 
has a low number density compared
with, for example, the Orion Nebula Cluster.  Due to its sparse
environment and young age, it is thought that very little dynamical
evolution has taken place  \citep{Kroupa03b}, and that the observed
stars are direct signatures of the star formation process in this
region  \citep{Luhman06}. For this reason, attempts have been made to
quantify the spatial distribution of stars and brown  dwarfs in Taurus
to test various formation hypotheses, including those that postulate a
different formation scenario  for brown dwarfs over stars
\citep[e.g.][]{Reipurth01,Thies07}.

In this paper we use the new minimum spanning tree (MST) method
\citep{Allison09a} to look for differences in the  distribution of
low- and high-mass objects in Taurus. We describe the observational
sample used in  Section~{\ref{observations}, before presenting the
  results in Section~\ref{results}. We compare the MST method  to
  other measures of spatial distribution in Taurus in
  Section~\ref{nearestN}, we discuss our results in Section~\ref{discuss} 
and we present our conclusions in Section~\ref{conclude}.

This is the first in a series of papers in which we will discuss the
formation of stars and brown dwarfs in Taurus  by considering the
process from prestellar cores to the subsequent effects of dynamical
evolution on  the cluster population. 
 
\section{The observational sample}
\label{observations}

\begin{figure*}
\begin{center}
\rotatebox{270}{\includegraphics[scale=0.75]{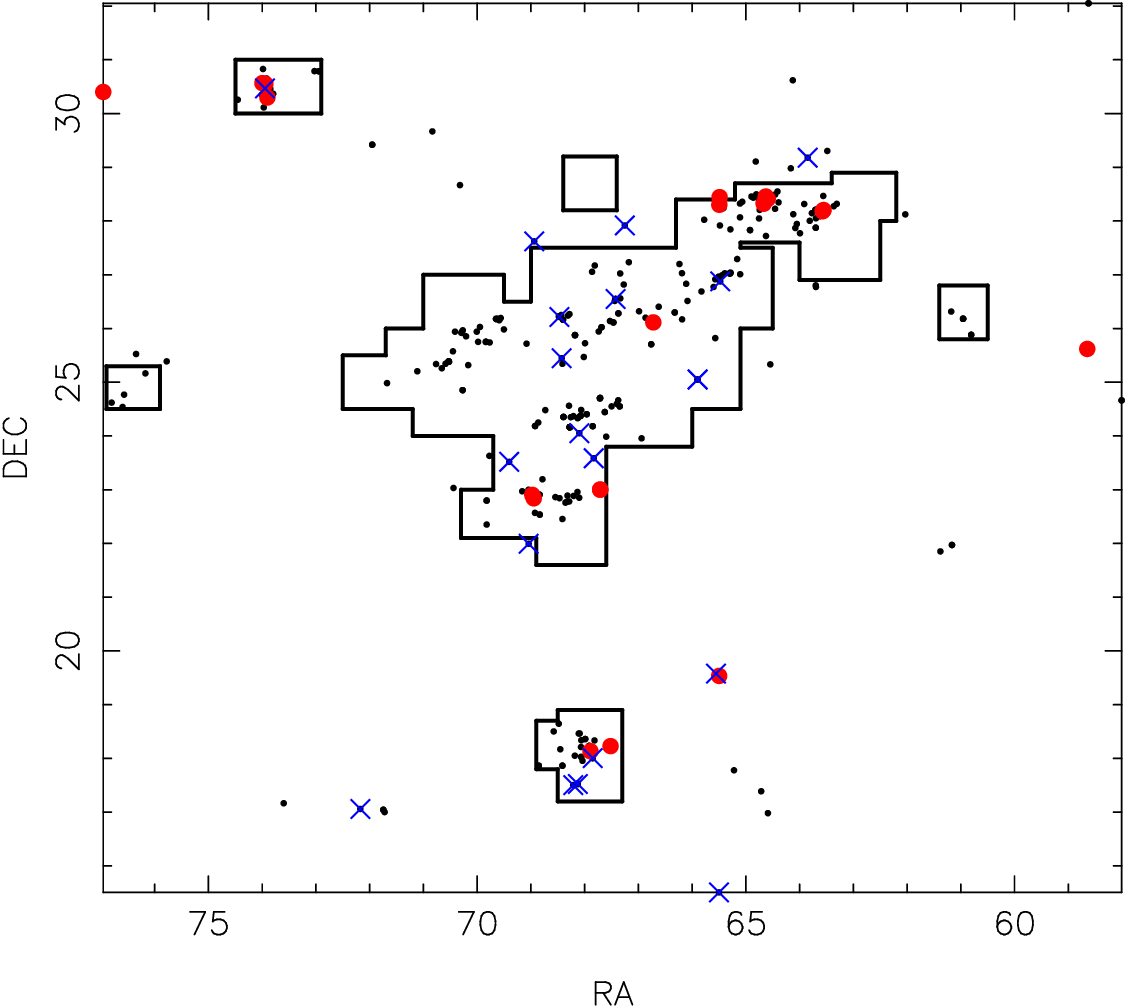}}
\end{center}
\caption[bf]{A map of the Taurus cluster showing the 361 objects in
  our dataset. The 20 least massive cluster  members are shown by the
  (blue) crosses and the 20 most massive cluster members are shown by
  the large (red) dots.  The areas of Taurus that are observationally
  complete \citep[surveys by][and references
    therein]{Briceno02,Luhman06,Guieu06,Luhman10}  are inside the solid lines.}
\label{Taurus_most_least}
\end{figure*}

Our primary database for the following analysis is a catalogue of 442
Taurus sources compiled by the XEST collaboration \citep{Guedel07}
as an ``input catalogue'' for the XEST project. This input catalogue
was compiled by cross-identifying objects between various previous
catalogues of Taurus members \citep[in particular from][]{Kenyon95,Briceno02,Palla02}
and general all-sky catalogues relevant for pre-main sequence stars.
Ancillary information such as photometric and
spectroscopic data, coordinates, masses and ages, was then extracted from
the individual catalogues, although for some all-sky  survey catalogues we
confined the search for counterparts to within a radius of 8 degrees of
the position RA(2000.0) = 4$^{\rm h}$ 25$^{\rm m}$ dec(2000.0) = 25 deg
(note that this constraint is irrelevant for the {\it identification} of
Taurus members which relies on previous, dedicated Taurus catalogues).
Information from SIMBAD and the 2MASS catalogues (essentially spectral types,
coordinates and photometry) was confined to the areas covered by the
XEST X-ray exposures (again, this does not affect the membership
identification relevant for our study). A condensed version
of the input catalogue for the areas covered by the XEST survey was
published in \citet{Guedel07} where the relevant catalogue
bibliography is also described.

Of these 442 catalogue sources, 293 have a mass
estimate derived from bolometric luminosity, $L_{\rm bol}$, and 
effective temperature, $T_{\rm eff}$, using
\citet{Siess00} isochrones with a relative uncertainty of order of 20
per cent \citep[see][]{Guedel07}.  Of the remaining 149 objects
without a mass listed in the catalogue, 55 have a known spectral
type. We used this spectral type to derive a mass estimate from
\citet{Siess00} isochrones assuming an age of 2 Myr. This yielded a
total of 328 Taurus members following the removal of
duplicates. Where available, binary companions were included in the
sample. Of these  328 objects, 20 do not appear in the more recent
compilation of Taurus members by \citet{Kenyon08} and we therefore
rejected them from the analysis. We will discuss  the possible effects
of hidden binaries and rogue non-members on our results in
Section~\ref{results}.

The XEST catalogue misses most of the recently discovered very low
mass stars and substellar members of the Taurus cloud. We therefore
completed the XEST sample with the low mass end of the Taurus
population taken from \citeauthor{Kenyon08}'s (\citeyear{Kenyon08})
compilation that lists 382 Taurus members. The latter database
includes 85 very low mass Taurus members not included in the XEST
database. Of these, only 53 have a spectral type listed in
\citet{Luhman10}. We used these spectral types to derive mass
estimates from \citeauthor{Siess00}'s (\citeyear{Siess00}) 2 Myr
isochrone.

Adding these more recent detections to the XEST source list eventually
yields a catalogue of 361 Taurus members with a mass estimate. We
conservatively estimate the relative error on the mass to be of order 
30 per cent. Alternatively, as a check to the robustness of our
results below, we also used \citeauthor{Luhman10}'s
(\citeyear{Luhman10}) list of 324 Taurus members with known spectral
types, for which we derived a mass estimate using
\citeauthor{Siess00}'s (\citeyear{Siess00}) 2 Myr isochrone.

We show a map of the Taurus cluster made with our data  in
Fig.~\ref{Taurus_most_least}. The (blue) crosses show the twenty least
massive objects (all are brown  dwarfs) in the cluster, whereas the
large (red) points show the twenty most massive stars in the cluster.
Extensive surveys of various areas of Taurus by
\citet{Briceno98,Briceno02,Luhman00,Luhman03,Luhman04,Luhman06,Guieu06}
are shown by the black outlines. It is thought that these areas are
more or less observationally complete, whereas the regions outside 
of these lines may not be  \citep{Luhman09,Luhman10,Monin10}.

\section{Results}
\label{results}

In this Section we describe the minimum spanning tree (MST) method
used to quantify mass segregation in  clusters before applying it to
sets of objects of similar mass in Taurus.

\subsection{The minimum spanning tree method}

Following \citet{Allison09a}, we adopt the minimum spanning tree (MST)
method to quantify the level of mass segregation  in Taurus. The MST
of a set of points is the path connecting all the points via the
shortest possible pathlength but which  contains no closed loops
\citep[e.g.][]{Prim57,Cartwright04}.

We use the algorithm of \citet{Prim57} to construct MSTs in our
dataset. We first make an ordered list of the separations  between all
possible pairs of stars\footnote{From this point onwards, when referring in 
general to `stars' in the cluster, we mean `stars \emph{and} brown dwarfs', 
as we are including all the objects in the observational
sample.}. Stars are then connected together in `nodes',
starting with the shortest separations and  proceeding through the
list in order of increasing separation, forming new nodes if the
formation of the node does not result in a closed loop.

\subsection{Quantifying mass segregation}

 Observationally, `mass segregation' is a term used to describe the
  central concentration of massive stars in a star cluster (the prime
  example probably being the Trapezium of massive stars at the centre
  of the Orion Nebula Cluster).  In addition, mass segregation is
  often used in dynamics to refer to the central concentration of
  massive stars, and the wider distribution of low-mass stars caused
  by energy equipartition due to two-body relaxation.

In this paper we will define `mass segregation' in terms of the
relative spatial distributions of stars in a particular
mass range with respect to other stars in a
cluster. This also allows us to define `inverse mass segregation' as an
under-concentration of a particular stellar mass range with  respect
to the other cluster members. Note that we can apply this definition
to low-mass stars/brown dwarfs, and by  describing a population of
low-mass stars as `inversely mass segregated' we do not mean that the
high-mass stars are  necessarily mass segregated.

We find the MST of the $N_{\rm MST}$ stars in the chosen subset and
compare this to the MST of sets of $N_{\rm MST}$ random  stars in the
cluster. If the length of the MST of the chosen subset is shorter than
the average length of the MSTs for the  random stars then the subset
has a more concentrated distribution and is said to be mass
segregated. Conversely, if the MST  length of the chosen subset is
longer than the average MST length, then the subset has a less
concentrated distribution, and is  said to be inversely mass
segregated. Alternatively, if the MST length of the chosen subset is
equal to the random MST length,  we can conclude that no mass
segregation is present.

By taking the ratio of the average random MST length to the subset MST
length, a quantitative measure of the degree of  mass segregation
(normal or inverse) can be obtained. We first determine the subset MST
length, $l_{\rm subset}$. We then  determine the average length of
sets of $N_{\rm MST}$ random stars each time; $\langle l_{\rm average}
\rangle$. There is a dispersion  associated with the average length of
random MSTs, which is roughly Gaussian and can be quantified as the
standard deviation  of the lengths  $\langle l_{\rm average} \rangle
\pm \sigma_{\rm average}$. However, we conservatively estimate the lower (upper) uncertainty 
as the MST length which lies 1/6 (5/6) of the way through an ordered list of all the random lengths (corresponding to a 66 per cent deviation from 
the median value, $\langle l_{\rm average} \rangle$). This determination 
prevents a single outlying object from heavily influencing the uncertainty. 
We can now define the `mass  segregation ratio' 
($\Lambda_{\rm MSR}$) as the ratio between the average random MST pathlength 
and that of a chosen subset, or mass range of objects:
\begin{equation}
\Lambda_{\rm MSR} = {\frac{\langle l_{\rm average} \rangle}{l_{\rm subset}}} ^{+ {\sigma_{\rm 5/6}}/{l_{\rm subset}}}_{- {\sigma_{\rm 1/6}}/{l_{\rm subset}}}.
\end{equation}
A $\Lambda_{\rm MSR}$ of $\sim$ 1 shows that the stars in the chosen
subset are distributed in the same way as all the other  stars,
whereas $\Lambda_{\rm MSR} > 1$ indicates mass segregation and
$\Lambda_{\rm MSR} < 1$ indicates inverse mass segregation,
i.e.\,\,the chosen subset is more sparsely distributed than the other stars.

As noted by \citet{Allison09a}, the MST method gives a quantitative
measure of mass segregation with an associated significance and   it
does not rely on defining the centre of a cluster (somewhat impossible
for a substructured region like Taurus). It also bypasses  the various
binning methods used in determining mass segregation through fitting a
density profile \citep[e.g.][]{Adams01,Littlefair03}  or tracing the
change in mass function with radius
\citep[e.g.][]{Gouliermis04,Sabbi08}. We
shall now apply the MST method to look for mass segregation in the
high-  and low-mass stellar (and substellar) populations in Taurus.

\subsection{High-mass cluster members}

In Fig.~\ref{Taurus_most_least} we show the location of the twenty
most massive objects in the cluster  ($m \gtrsim$\,1.2\,M$_\odot$) by
the large (red) points. Several are within the central aggregates, but
others are located in both the northern and southern Gomez groups 
\citep{Gomez93}. In Fig.~\ref{Taurus_hm_MST} we show $\Lambda_{\rm MSR}$ as 
a function of the number of stars in an MST for the highest mass 
stars. $\Lambda_{\rm MSR} = 1$, indicating no difference between the
distribution of these stars and other stars, is
shown by the dashed line. 

\begin{figure}
\begin{center}
\includegraphics[scale=0.6]{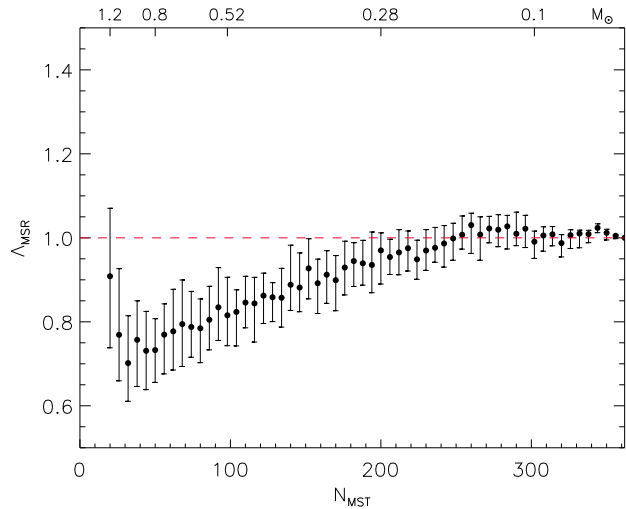}
\end{center}
\caption[bf]{The evolution of the mass segregation ratio,
  $\Lambda_{\rm MSR}$, with respect to the $N_{\rm MST}$  most massive
  stars in Taurus. Error bars show the 1/6 and 5/6 percentile values from 
the median, as described in the text. The dashed line indicates $\Lambda_{\rm MSR} = 1$,
  i.e.\,\,no mass segregation. We also show the lowest mass within 
 $N_{\rm MST}$ stars on the top axis.}
\label{Taurus_hm_MST}
\end{figure}

Inspection of Fig.~\ref{Taurus_hm_MST} shows that the highest mass
stars in the cluster are spread more widely than other stars, ie. they
are inversely mass segregated,  with a 
trough at $\Lambda_{\rm MSR} = 0.70 \pm 0.10$. 

\subsection{Low-mass (brown dwarf) cluster members}

In Fig.~\ref{Taurus_most_least} we show the location of the twenty
least massive objects (all of which are brown dwarfs) by  the (blue)
crosses. Most are concentrated in the central aggregates, but there
are several in the outlying clumps. We show the calculation of
$\Lambda_{\rm MSR}$ as a function of the number of stars in an MST 
for the low-mass objects in Fig.~\ref{Taurus_lm_MST}. Again, 
$\Lambda_{\rm MSR} = 1$, indicating no mass segregation, is shown
by the dashed line. 

\begin{figure}
\begin{center}
\includegraphics[scale=0.6]{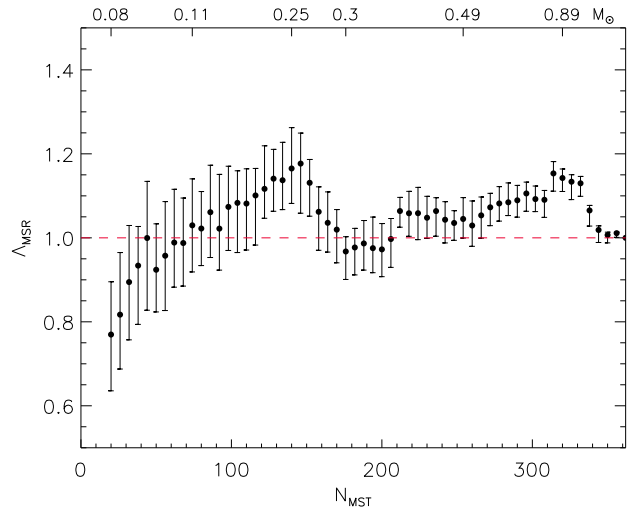}
\end{center}
\caption[bf]{The evolution of the mass segregation ratio,
  $\Lambda_{\rm MSR}$, with respect to the $N_{\rm MST}$  least
  massive stars (brown dwarfs) in Taurus. Error bars show the 1/6 and 
5/6 percentile values from the median, as described in the text. The dashed line indicates
  $\Lambda_{\rm MSR} = 1$, i.e.\,\,no mass segregation. We show the 
  highest mass within $N_{\rm MST}$ stars on the top axis. }
\label{Taurus_lm_MST}
\end{figure}

Fig.~\ref{Taurus_lm_MST} shows that the distribution of brown
dwarfs in the cluster is roughly uniform, 
fluctuating around $\Lambda_{\rm MSR} = 1$, with no
clear trend towards either mass segregation, or inverse mass
segregation. There are hints that the low-intermediate mass stars may
be mass segregated  (see Section~\ref{allmembers}), and the brown
dwarfs have $\Lambda_{\rm MSR} < 1$, but overall the plot is consistent with
there being no difference between the distribution of low-mass objects
and other objects.

\subsection{MSTs for all cluster members}
\label{allmembers}

In a new variation of the MST method, we calculate the MSTs for stars
as a function of mass. This is achieved by taking  the MST of a subset
of the $N_{\rm MST}$ lowest-mass objects (we take the average of
$N_{\rm MST} = 40$ objects, rather than  $N_{\rm MST} = 20$, to reduce
the uncertainties), and then sliding through the mass range in steps
of 10 objects. For example,  the first subset contains the 40
lowest-mass objects, the second contains the 10 -- 50 lowest mass
objects, the third contains  the 20 -- 60 lowest mass objects and so
on\footnote{Because we have 361 objects in our sample, the final
  subset contains 41 stars,  rather than 40.}. We then calculate
$\Lambda_{\rm MSR}$ as before, and plot it in Fig.~\ref{slide_MST_TMC}
as a function of the  highest mass object in each subset. 

It should be  noted that in this method the data points are not 
independent of one another, with
  each data point including some of the same information as those in
  the two bins either side. However, if we move through the dataset in steps 
of 30 objects, without any overlap, and compare the MST of each subset to 
random MSTs of 30 objects, the main results still hold.

In Fig.~\ref{slide_MST_TMC} we see again that, when compared to the MST
of random subsets of objects, the brown dwarfs have a  mass
segregation ratio consistent with unity. Stars with masses in the
range $0.1 - 0.25$\,M$_\odot$ appear to be slightly more concentrated 
(mass segregated), as do stars in the range $0.45 -
0.8$\,M$_\odot$. In both mass regimes, the  mass segregation ratio is
$\Lambda_{\rm MSR} = 1.25 \pm 0.15$.

\begin{figure}
\begin{center}
\rotatebox{270}{\includegraphics[scale=0.33]{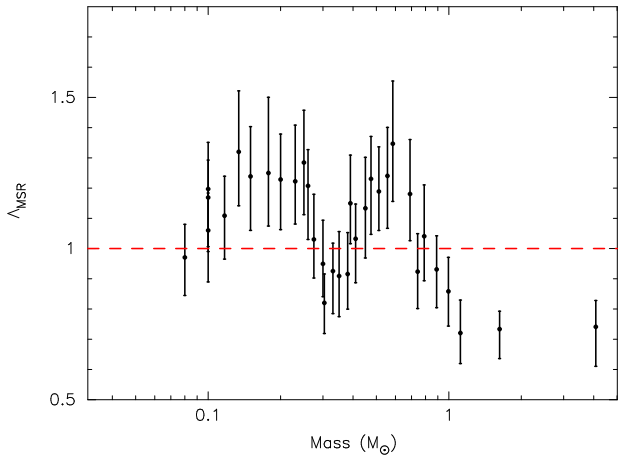}}
\end{center}
\caption[bf]{The evolution of the mass segregation ratio,
  $\Lambda_{\rm MSR}$, as a function of mass for subsets of 40
  stars. We plot  the highest mass object in each subset. The dashed
  line indicates $\Lambda_{\rm MSR} = 1$, i.e.\,\,no mass
  segregation.}
\label{slide_MST_TMC}
\end{figure}

Interestingly, stars with masses
centred on $0.3$\,M$_\odot$ appear to have a wider distribution (slightly inversely mass
segregated), with a trough at $\Lambda_{\rm MSR} = 0.80 \pm 0.10$. In Fig.~\ref{position0p3} 
we show the location of stars with mass in the range  0.25 -- 0.35\,M$_\odot$ 
by the plus signs. Most of the stars in our sample with this mass have 
spectral types in the range M2 -- M6, in the regime in which the observations 
may be incomplete {\em outside} the clumpy regions of the
cluster \citep{Guieu06,Luhman06}. 

\begin{figure}
\begin{center}
\rotatebox{270}{\includegraphics[scale=0.45]{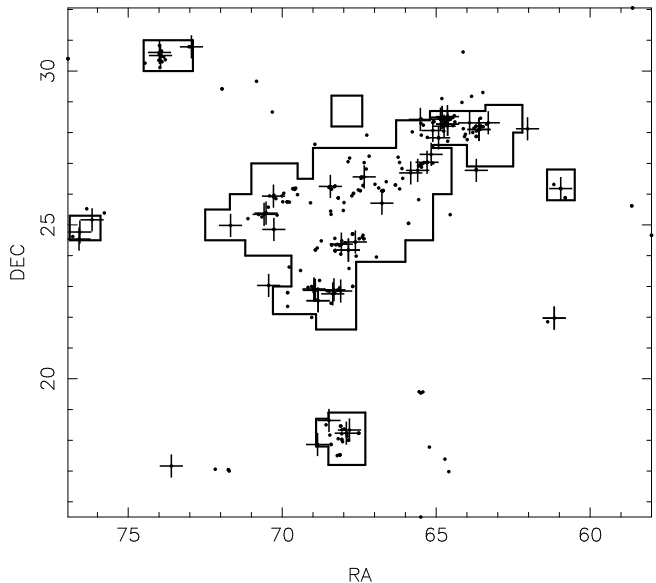}}
\end{center}
\caption[bf]{A map of the Taurus cluster, with the positions of stars of mass 0.25 -- 0.35\,M$_\odot$ 
shown by the plus signs. The deep fields surveyed by  \citet{Briceno02,Luhman06,Guieu06}, 
and references therein, are inside the solid lines.}
\label{position0p3}
\end{figure}

Our result implies that if there is a deficiency of M2 -- M6
objects, then those that are missing  should be located within the
clumps, assuming that the anomalous $\Lambda_{\rm MSR}$ around $0.3$\,M$_\odot$ is a 
real feature, and that these objects  do not form via a different
mechanism to e.g.\,\,objects of mass 0.2 and 0.5\,M$_\odot$.

Above a mass of $\sim 0.9$\,M$_\odot$, the stars in each subset are
inversely mass segregated with respect to random stars in the cluster,
confirming the results shown in Fig.~\ref{Taurus_hm_MST}. The level of
inverse mass segregation reaches a minimum value of $\Lambda_{\rm MSR} =
0.70 \pm 0.10$.  Whilst this can be said to be a rather modest level of inverse mass
segregation, it is markedly different to the MSR for stars with masses
of  $\sim 0.5$\,M$_\odot$.

\subsection{Potential uncertainties} 

In this section we briefly discuss the caveats associated with our
results, namely the main observational uncertainties that would
affect the resultant $\Lambda_{\rm MSR}$ values.

\subsubsection{Mass determination}

The mass determinations for most objects in our observational sample
are likely to be uncertain by up to 30 per cent. It is not possible to
directly quantify this in the determination of $\Lambda_{\rm MSR}$, as
this value is obtained by calculating pathlengths between objects, and
is not weighted by the object's mass\footnote{An advantage of the MST
  method is that it does not require an absolute mass
  determination. One can use e.g.\,\,absolute magnitude
  \citep{Sana10}.}. In order to estimate the effect of the mass
uncertainty on  our result, we randomly added or subtracted up to 30
per cent of the mass from each object, and then performed our analysis
on  this data. From multiple realisations of this experiment, we find
no significant difference to the main result that the most massive
stars  are inversely mass segregated, and the low-mass stars are
slightly mass segregated.  However, the inverse mass segregation of
objects at 0.3\,M$_\odot$ is largely erased each time,  due to the
addition of random noise to the mass of each star. The effect of this
process is place the stars that show strong segregation into different
bins,  diluting the result. 

\subsubsection{Binary companions}

We include objects that were listed as binary systems in the catalogue
of \citet{Guedel07} in our analysis. In order to test for  the effects
of close, or hidden binaries that may be missing from our data, we
performed two experiments on the data. Firstly, if  an object was
multiple, we removed it, and its companion(s) from the dataset
altogether. This does not alter the the results in  any way. Secondly,
we summed the masses of the components and added these to the primary,
thereby accounting for (and probably overestimating) the effects of
hidden companions on the mass. Again, negligible differences  to the
main results were found.

\subsubsection{Rogue cluster members}

By comparing the XEST catalogue of \citet{Guedel07} with that in
\citet{Kenyon08}, we are confident that there are no non-members
masquerading in our dataset. However, should there be any rogue
members in our sample, they will affect the analysis in 2-D only;
i.e. background/foreground field stars will not cause an MST length to
be overly long in the third dimension.  Field stars in the  diffuse
regions (outside of the black outline in Fig.~\ref{Taurus_most_least})
could adversely affect the results, but we suggest that  the chances
of this are minimal for two reasons. Firstly, the MST results are
identical whether we include or exclude the 20 members  of our sample
not found in the catalogue of \citet{Kenyon08}. Secondly, using the
largely independent sample from \citet{Luhman10},  we also find very
similar MST results (see Section~\ref{nearestN}). This suggests that
our observational sample would have to change drastically (and that
there  would have to be a significant number of rogue stars
distributed differently to the cluster members) before the MST results
are  adversely compromised.

\subsubsection{Missing B-type stars}

The Initial Mass Function (IMF) in Taurus has been the subject of much
debate.  Initially, it was thought that Taurus was deficient in both
brown dwarfs  \citep{Briceno02} and high mass stars
\citep{Walter91}. This contravenes  the universality of the IMF, which
appears the same in most star forming  regions
\citep*{Kroupa02,Bastian10}. Recently, the discovery of many brown
dwarfs  \citep{Luhman04,Guieu06} has removed the deficit in the
low-mass regime.

However, if one extrapolates the IMF to the high-mass regime, there
could be  up to 40 B-type stars ``missing'' from Taurus
\citep{Walter91}. \citet{Walter91}  proposed that 21 stars in the
Cas-Tau OB association were related to Taurus. However,  10 of these
candidate members lie outside the field of view in
Fig.~\ref{Taurus_most_least},  and presumably have low-mass stars
associated with them, for which we have no information.  To determine
the effect of these stars on our results, we first added all 21
candidates to  our object list, before running the MST on this, and a
list containing only the 11 stars  that lie within our field of
view. In both cases, the net result is the B-type stars are  even more
inversely mass segregated than solar-mass stars.

In short, if there are missing B-type stars from our observational
sample, we would expect them to simply reinforce our  main
results. However, we note that in some cases, sampling an IMF to
populate a low number  cluster such as Taurus could in principle lead
to a deficiency in a particular mass of  object \citep{Parker07}.

\subsubsection{Incompleteness in the low-mass regime}

In Fig.~\ref{Taurus_most_least} the fields for which the observations
are  thought to be entirely complete \citep[][and references
  therein]{Luhman10,Monin10}  are indicated by the solid
lines. Outside these regions, it is possible that  surveys of Taurus
may have missed objects, particularly low-mass stars and brown
dwarfs. Such missing objects may impact upon the results of our MST
technique.  To qualify the potential effects of missing objects, we
have run the MST on the  central region only (encompassed by the solid
line in Fig.~\ref{Taurus_most_least}).  The results are shown in
  Fig.~\ref{Taurus_hm_central}. The most massive objects have a  mass
  segregation ratio $\Lambda_{\rm MSR} = 0.81^{+0.10}_{-0.05}$, which
  is not as  extreme a trough as the $\Lambda_{\rm MSR} = 0.70 \pm
  0.10$ found for the whole association.  However, in
  Fig.~\ref{Taurus_most_least}  one can clearly see that many of the
  most massive  stars in the association are located outside of the
  central region. If the sparsely populated regions in between the 
  central region and the groups \emph{are} more or less complete, then omitting 
  the outlying regions from the analysis is potentially adding a bias to the results because 
  we are no longer considering the entire star forming region.  

Interestingly, \citet{Kirk10}  recently studied the sub-groups of
stars within Taurus and found that the most massive stars in  the
groups are mass segregated. \citet{Kirk10} determined the centre of each sub-group, 
and then calculated the offset from the centre for each star. They find that the most massive star 
in each group has an offset which is significantly lower than the median. We also reproduce this result if we
calculate   $\Lambda_{\rm MSR}$ on, for example the L1551 group \citep{Gomez93} 
enclosed by the black outline at the  bottom of
Fig.~\ref{Taurus_most_least}. For each of the subgroups we find that
the most massive  stars are mass segregated. However, the calculated
values for $\Lambda_{\rm MSR}$ are not as   robust as those for the
whole association, due to low-number statistics. We prefer to consider
the  entire association in our analysis, as the sub-groups are
interlinked via gas filaments, so star  formation must be happening on
a global scale. 

Additionally, a great  deal of observational effort
has gone into improving the completeness of the entire association
and it may be that the data are more or less complete \citep[e.g.][]{Guieu06,Luhman10}.  
Furthermore, star formation is recognised to be more prominent in filaments, which have 
strong CO signatures and high dust extinction \citep[e.g.][and references therein]{Palla02,Schmalzl10}. Indeed, 
\citet{Luhman09} show that dust extinction \citep{Dobashi05} is strongly correlated with known 
 members, with little evidence of filaments elsewhere, suggesting that the 
stellar census is complete.

Finally, we note that any theory of star formation in Taurus must reproduce the inverse
mass  segregation we observe over a large scale, \emph{and} the
localised mass segregation observed  by \citet{Kirk10}. We will
examine this in detail in a forthcoming paper.

\begin{figure}
\begin{center}
{\includegraphics[scale=0.6]{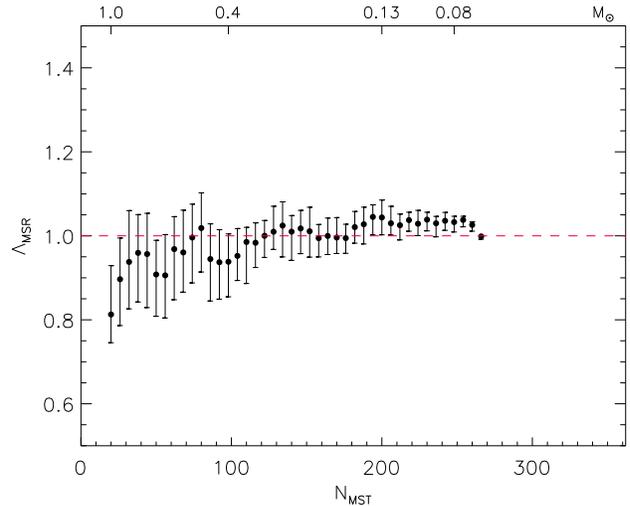}}
\end{center}
\caption[bf]{As Fig.~\ref{Taurus_hm_MST}, but for the central region 
marked by the black outline in Fig.~\ref{Taurus_most_least}. The dashed line indicates 
$\Lambda_{\rm MSR} = 1$,  i.e.\,\,no mass segregation. We also show the lowest mass within 
 $N_{\rm MST}$ stars on the top axis.}
\label{Taurus_hm_central}
\end{figure}

\subsubsection{Extinction}

A further, related issue to completeness is the variation of extinction across the cluster. 
In the most clustered regions, the faintest objects may not be detected due to their being 
embedded in the gas. As a check, we discarded all objects with an extinction A$_{\rm v} > 4$ and repeated 
the analysis. Again, we find no discernible difference to the results.

\section{Comparison with other methods and datasets}
\label{nearestN}

In this section we compare the results of the MST analysis of our
Taurus dataset with other datasets and with other methods that
have previously been used to analyse the spatial distribution of objects
in Taurus.

\subsection{Comparison with other data}

In recent work, \citet{Luhman10} provided a list of 324 members of
Taurus for which spectral types could be assigned to each object.
From these spectral types, masses were inferred using the isochrones
of \citet{Siess00}. As an independent test of our method, we  repeat
the step MST analysis in Section~\ref{allmembers} for the objects in
\citeauthor{Luhman10}'s sample and our results are shown in
Fig.~\ref{slide_MST_luhman2010}.  It should be noted that the subsets
of objects lie in slightly different locations to those calculated
using our dataset in Fig.~\ref{slide_MST_TMC}, due to the fact that
there are 37 fewer  members overall, and objects with similar spectral
types are assigned the same masses, causing the `pile-up' of mass
segregation ratios at some  mass values. However, in general, the
results are very similar to those using our data; the brown dwarfs
have  $\Lambda_{\rm MSR} \sim 1$, whereas  stars with masses less than
1\,M$_\odot$ appear mass segregated, with the anomalous feature still
prevalent at 0.3\,M$_\odot$. The data from  \citet{Luhman10} are also
consistent with  $\Lambda_{\rm MSR}$ = 0.7 (within the uncertainties)
for the most massive objects in Taurus. 

\begin{figure}
\begin{center}
\rotatebox{270}{\includegraphics[scale=0.33]{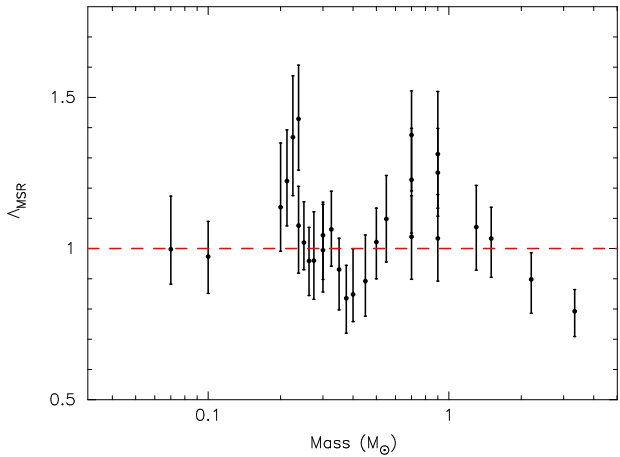}}
\end{center}
\caption[bf]{As Fig.~\ref{slide_MST_TMC}, but computed with data
  provided in \citet{Luhman10}. The mass segregation ratio,
  $\Lambda_{\rm MSR}$, is plotted as a function of the most massive
  object in each subset of 40 stars. The dashed line indicates
  $\Lambda_{\rm MSR} = 1$, i.e.\,\,no mass segregation.}
\label{slide_MST_luhman2010}
\end{figure}

\subsection{The $\mathcal{R}_{\rm ss}$ ratio of substellar-stellar objects}

Previous studies into the spatial distribution of brown dwarfs in
Taurus measured the ratio of brown dwarfs to  stars for both the whole
cluster and the separate aggregates:
\begin{equation}
\mathcal{R}_{\rm ss} = \frac{N(0.02 < m/{\rm M}_\odot \leq
  0.08)}{N(0.08 < m/{\rm M}_\odot \leq 10)}.
\label{Rss}
\end{equation}
This ratio has been calculated for the whole Taurus association
\citep{Briceno02,Luhman04,Guieu06}, resulting in a range  of values
depending on the chosen dataset. For example, \citet{Briceno02} find
$\mathcal{R}_{\rm ss} = 0.13 \pm 0.04$,  \citet{Luhman04} finds
$\mathcal{R}_{\rm ss} = 0.18 \pm 0.04$ and \citet{Guieu06} find
$\mathcal{R}_{\rm ss} = 0.23 \pm 0.04$.  \citet{Guieu06} also applied
the $\mathcal{R}_{\rm ss}$ ratio to the various aggregates and
concluded that the brown dwarfs  are less abundant (by a factor of
$\sim$~2) compared to stars in the aggregates than for the overall
cluster.

An overall cluster value of $\mathcal{R}_{\rm ss} = 0.23 \pm 0.04$ is
consistent with the Trapezium cluster \citep{Briceno02},  whereas
lower values suggest a deficiency in the substellar IMF. However,
\citet{Luhman06} argues that the  $\mathcal{R}_{\rm ss}$ ratio is
strongly biased by the assignment of spectral type to a particular
object (as this changes  both the numerator and denominator of
Eqn.~\ref{Rss}). 

A further, related problem lies in determining the completeness of the
substellar population. For example,  if we have 30 brown dwarfs and
220 stars, $\mathcal{R}_{\rm ss} = 0.14$. If a further 10 brown dwarfs
are added to the sample,  the ratio of substellar to stellar objects
becomes $\mathcal{R}_{\rm ss} = 0.18$. In other words, a normal IMF
can appear  abnormal simply due to observational incompleteness. Such
a change to the sample would not drastically affect the results of the
MST technique, unless the majority of the missing brown dwarfs were
spatially distributed in a very different fashion to  other objects of
similar mass in the sample.

\subsection{Nearest neighbour distances}

\begin{figure*}
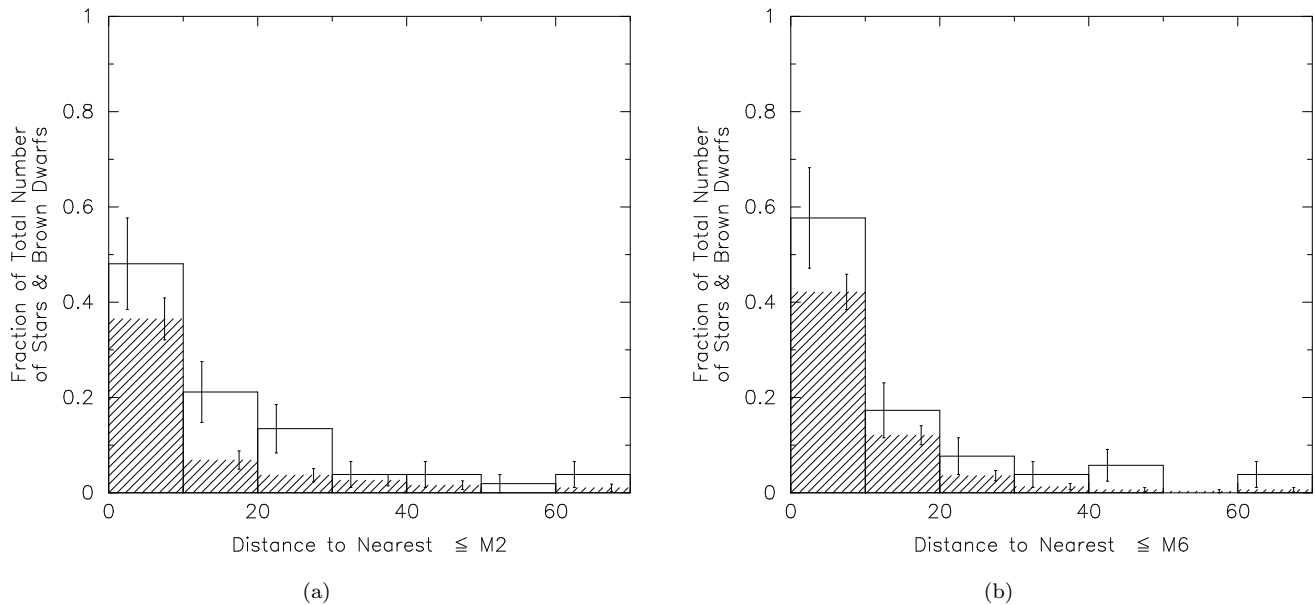

  \begin{center}
\setlength{\subfigcapskip}{10pt}
\subfigure[]{\label{fmult-a}\rotatebox{270}{\includegraphics[scale=0.4]{Luhman_M6-M2_NOBINS_2.ps}}}
\hspace*{0.6cm}
\subfigure[]{\label{fmult-b}\rotatebox{270}{\includegraphics[scale=0.4]{Luhman_M6-M6_NOBINS_2.ps}}}
  \end{center}
  \caption[bf]{The distances to nearest neighbours of stars and brown
    dwarfs. In (a) we show a distribution of the  distances to the
    nearest $\leq$~M2 star from: (i) a $>$~M6 brown dwarf (the open
    histogram with error bars on the left  of each bin); and (ii) a
    $\leq$~M2 star (the hashed histogram with error bars on the right
    of each bin). Each histogram  is normalised to the total number of
    $>$~M6 or $\leq$~M2 objects. In (b) we show a distribution of the
    distances to  the nearest $\leq$~M6 star from: (i) a $>$~M6 brown
    dwarf (the open histogram with error bars on the left of each
    bin);  and (ii) a $\leq$~M6 star (the hashed histogram with error
    bars on the right of each bin). Each histogram is normalised  to
    the total number of $>$~M6 or $\leq$~M6 objects.}
  \label{Luhman_nn}
\end{figure*}

In order to minimise the perceived biases associated with the
$\mathcal{R}_{\rm ss}$ ratio, \citet{Luhman06} adopted the nearest
neighbour distance as a method of quantifying the spatial distribution
of brown dwarfs in Taurus. In his analysis,  \citet{Luhman06}
classified objects with spectral type $>$~M6 as brown dwarfs, and
objects $\leq$~M6 as stellar objects.  To account for the potential
incompleteness in the range M2 -- M6 \citet{Luhman06} also made a
sub-classification of stars  as $\leq$~M2, and compared objects with
$>$~M6 to both $\leq$~M6 and $\leq$~M2. 

For each object class, \citet{Luhman06} determined the distance to the
nearest neighbour. He examined the distance from  each $>$~M6 (brown
dwarf) and $\leq$~M2 (star) to the nearest $\leq$~M2; and the distance
from each $>$~M6 (brown dwarf)  and $\leq$~M6 (star -- second
definition) to the nearest $\leq$~M6 -- see his fig.~14. We repeat his
analysis for the  dataset used here and our results are shown in
Fig.~\ref{Luhman_nn}. 

We agree with the conclusion of \citet{Luhman06}; the distances
between brown dwarfs and stars, and stars and stars, do  not differ by
much in our dataset. However, there do appear to be subtle variations
in the spatial distribution as a  function of the mass of the object
in Taurus (recall
Figs.~\ref{slide_MST_TMC}~and~\ref{slide_MST_luhman2010}). These
differences are not apparent in the nearest neighbour analysis. In
Fig.~\ref{Luhman_nn}, the distributions of nearest  neighbour
distances between any chosen groups of objects are
identical. \citet{Guieu06} find a similar result, and  both authors
found the distribution of stellar and substellar nearest neighbour
distances to be consistent.

We therefore caution against using the mean nearest neighbour distance
to define the spatial distribution of brown dwarfs  compared to stars
in a cluster. If the mass function of Taurus is normal, we would
expect there to be 4 -- 5 times as  many stars as brown dwarfs in the
cluster \citep{Andersen08}. If we calculate the average nearest
neighbour distance  between the brown dwarfs in our sample, we obtain
a value of 33 arcminutes, compared to a value of 11 arcminutes between
stars. However, this technique is biased towards obtaining smaller
nearest neighbour distances for stars because there are more of these
objects in the cluster than brown dwarfs. Therefore, the stars are
more likely to be closer to other stars than the brown  dwarfs are to
brown dwarfs. 

If we compare the MST length between brown dwarfs to the MST length of
random sets of stars, we obtain a (largely) unbiased  determination of
the spatial distribution of these objects, and we are also able to
pick out the subtle differences in  the distribution of intermediate
mass stars and the highest mass stars (see Figs.~\ref{slide_MST_TMC}~and~\ref{slide_MST_luhman2010}).

Finally, we note that other comparisons between the MST technique and
nearest neighbour distance also find the MST to be  a more robust
determination of spatial distribution \citep{Gutermuth09}.

\section{Discussion}
\label{discuss}

We have calculated $\Lambda_{\rm MSR}$ \citep{Allison09a} for stellar
and substellar objects  across the entire Taurus association. We
find that the most massive stars in  the cluster ($m >
1.2$\,M$_\odot$) are slightly inversely mass segregated with respect
to random stars, with a trough at  $\Lambda_{\rm MSR} = 0.70 \pm 0.10$
($\Lambda_{\rm MSR} = 1$ indicates no mass segregation). This result
is unusual in that Orion (often considered to be  a `typical' star
cluster) displays mass segregation  of the most massive cluster
members (independent of the method used to define mass segregation),
with little or no mass segregation below 5\,M$_\odot$
\citep{Allison09a}. Currently, the only other cluster to have been
analysed using  the MST method is Trumpler~14, and this cluster is
similar to Orion in that it displays prominent mass segregation of the
most massive  stars \citep[$> 10$\,M$_\odot$,][]{Sana10}.

If the data are complete, they suggest that brown dwarfs are
distributed in a slightly different way to most low-mass  stars,
although within the uncertainties the two distributions are fairly
similar. However, if brown dwarfs form via a different  mechanism to
low-mass stars \citep[e.g.][]{Reipurth01,Thies07} then the observed
difference may be real  \citep[however see e.g.][for arguments that
  their formation is similar to that of low-mass hydrogen burning
  stars]
{Padoan02,Padoan04,Whitworth07,Stamatellos07b,Bate09,Whitworth10}.

Taking the results of this study at face value lead to several
conclusions.
\begin{itemize}

\item Firstly, the highest-mass stars in Taurus ($m > 1.2$\,M$_\odot$)
  are more widely distributed than average.  

\item Secondly, that brown dwarfs and very low-mass stars ($m <
  0.15$\,M$_\odot$)  are distributed randomly in the cluster and are
  not found preferentially either within or outside clumps.  

\item Thirdly, that intermediate-mass stars ($0.15 < m/$M$_\odot <
  0.7$) are more concentrated than a random selection of stars.  

\item Finally, stars of $\sim 0.3$\,M$_\odot$ are an exception to the
  concentration of intermediate-mass stars, seemingly significantly
  more widely distributed than stars of even slightly higher or lower
  masses.

\end{itemize}

A visual inspection of
Figs.~\ref{Taurus_most_least}~and~\ref{position0p3} does suggest that
the first three conclusions are at least plausible, especially that
the most massive stars are more sparsely distributed.  However, the
finding that stars of $\sim 0.3$M$_\odot$ are more sparsely
distributed  than stars slightly more or less massive ($\Lambda_{\rm
  MSR} = 0.8$ compared to 1.25) is rather odd and we will return to
this later.

Taurus is dynamically young and relatively unevolved.  The stellar and
gas densities are closely related \citep{Gomez93,Monin10},  and stars
are still forming with at least 20 prestellar cores found in the
cluster \citep{Kirk05}.  Therefore, at least to some extent, the
current positions of the stars follow where they formed.  That
higher-mass stars are found preferentially isolated compared to
intermediate-mass stars suggests that they formed in different places.
This may reflect how cores fragment, or possibly how their masses are
distributed.  It may be that cores that are close together fragment
more, forming groups of intermediate-mass stars whilst more isolated
cores tend to form fewer, but larger, stars.  Alternatively,  perhaps
each core only produces one or two objects, but that lower-mass cores
cluster more.  (It may be argued that these are two are equivalent.)
We note that the fragmentation scenario should  also produce the
  localised mass segregation of the sub-groups in Taurus,  as found by
  \citet{Kirk10}.

Brown dwarfs may be distributed differently to {\em all} stars of
whatever mass.  The statistical significance of this result is too
poor to draw any firm conclusions as the total sample size in Taurus
is rather small.  But this may suggest that brown dwarfs form as a
different population to stars in some way (or that very low-mass cores
are distributed differently).  Strong ejections
\citep[e.g.][]{Reipurth01}  would be expected to provide a fairly
strong signature of inverse mass segregation \citep[as dynamics would
  not have enough time to erase much of the signature,][]{Goodwin05c}
and so can probably be excluded as also found by \citet{Luhman06}; see
also \citet{Joergens06}.  That brown dwarfs are not found to be
associated with higher-mass stars suggests that disc fragmentation
around larger stars is not the formation mechanism behind most brown
dwarfs in Taurus \citep{Stamatellos07b}.   We note that gentle
liberation from binaries may give a slightly sparser  distribution of
brown dwarfs when compared to low-mass stars \citep{Goodwin07b}.

It would seem unlikely that stars of $0.3$\,M$_\odot$ would form or
dynamically evolve in a significantly different way to stars of mass
$0.2$\,M$_\odot$ or $0.4$\,M$_\odot$.  It is far more plausible that
this effect is due to incompleteness, or errors in the mass
determinations  of these objects.  Indeed, the spectral types that are
missing, M2 -- M6, may be incomplete \citep{Guieu06,Luhman06}  {\em
  outside} the clumpy regions of the cluster.  However, for this
result to be an artifact of incompleteness this particular spectral
range must be incomplete {\em inside} the clumps; more M2 -- M6 stars
away from clumpy regions will make the effect more extreme and not
less.  For this result to be due to incompleteness there must be
either (a) more $<$M2 and $>$M6 stars in sparser regions to lengthen
the MSTs of these types and to lengthen the average MSTs, or (b) more
stars of M2 -- M6 within the clumps.

Finally, we note that if the masses of all objects in Taurus were
subject to non-systematically change by  up to 30 per cent, then the
feature at $0.3$\,M$_\odot$ may disappear. Further  work to better
constrain the masses of these objects would obviously be desirable. 

We will return to a more detailed theoretical analysis of these
results in a future paper.

\section{Conclusions}
\label{conclude}

We have applied the minimum spanning tree (MST) method \citep{Allison09a} 
to search for mass segregation (both normal and inverse) in the stellar 
and substellar populations of the Taurus association. To this end, 
we determine the MST length of the 20 least massive stars and compare 
this with the MST lengths of random sets of stars. We repeat the procedure 
for the MST length of the 20 most massive stars. The level of mass 
segregation is then quantified via the mass segregation ratio ($\Lambda_{\rm MSR}$, 
where $\Lambda_{\rm MSR}$ = 1 corresponds to no mass segregation). 

We also apply a new variation of the MST method to compare the MST 
lengths of subsets of 40 objects to 40 random objects, thereby 
allowing us to trace the evolution of $\Lambda_{\rm MSR}$ as a function of object 
mass. This enables the mass segregation ratio of intermediate-mass 
objects to be calculated.

We determine $\Lambda_{\rm MSR}$  for the most massive stars ($m
\gtrsim$\,1.2\,M$_\odot$) in Taurus and find them to be slightly
inversely mass segregated ($\Lambda_{\rm MSR} = 0.70 \pm 0.10$),
i.e.\,\,preferentially located towards the outskirts of  the
cluster. This is unusual in that other star clusters show mass
segregation of the most massive stars  \citep{Allison09a,Sana10},
although such clusters are more massive, and dense, than Taurus.

We find that the brown dwarfs in Taurus have a mass segregation ratio consistent with no
mass segregation, although we find tentative evidence  that intermediate 
mass stars ($0.15 < m/$M$_\odot < 0.7$) show slight mass segregation, 
with $\Lambda_{\rm MSR} = 1.25 \pm 0.15$. 

These results suggest that brown dwarfs are distributed randomly
in the cluster, whilst intermediate-mass stars are generally
concentrated in clumpy regions, and higher-mass stars are distributed
more widely than average.  We note that the observations of stellar and
substellar objects in Taurus may be incomplete for spectral types
later than M2, and further surveys are desirable  in order to
determine whether low-mass stars are distributed differently to brown
dwarfs. Whilst incompleteness, especially away from the populous
well-studied regions may effect our conclusions for low-mass stars,
it is unlikely that any higher-mass stars are missing from
the surveys of Taurus and so, unless there is a significant population
of low-mass stars away from the known clumps, this result is robust.

Our method avoids the need for the sometimes arbitrary choice of
cluster centre necessary in radially-dependent searches  for mass
segregation. It also directly compares the path length between objects
of similar mass and random objects, rather  than the nearest neighbour
distance between stars and brown dwarfs, or the number ratio of brown
dwarfs to stars in a  particular region and we consider it to be a
more quantitative measure of mass segregation than previous
techniques.  In a follow-up paper, we will use the MST method to
compare models of prestellar core fragmentation with  the
observational data (Parker et al. in prep).

\section*{Acknowledgements}

RJP thanks Vik Dhillon for enabling the majority of this work to be undertaken 
at Sheffield. We thank the referee, Cathie Clarke, for helpful comments which greatly improved 
the original text. RJP thanks Helen Kirk for interesting discussions, and for making the results of 
her work available to us prior to publication. RJP and RJA acknowledge financial support from 
 STFC. RJP, JB, SPG, EM  and RJA acknowledge financial support from the EU Research 
 Training Network ``CONSTELLATION''. 

\bibliographystyle{mn2e}
\bibliography{taurus_ref}

\begin{thebibliography}{}

\bibitem[\protect\citeauthoryear{Adams, Stauffer, Monet, Skrutskie \&
  Beichman}{Adams et~al.}{2001}]{Adams01}
Adams J.~D.,  Stauffer J.~R.,  Monet D.~G.,  Skrutskie M.~F.,    Beichman
  C.~A.,  2001, AJ, 121, 2053

\bibitem[\protect\citeauthoryear{Allison, Goodwin, Parker, {Portegies Zwart},
  de Grijs \& Kouwenhoven}{Allison et~al.}{2009}]{Allison09a}
Allison R.~J.,  Goodwin S.~P.,  Parker R.~J.,  {Portegies Zwart} S.~F.,  de
  Grijs R.,    Kouwenhoven M. B.~N.,  2009, MNRAS, 395, 1449

\bibitem[\protect\citeauthoryear{Andersen, Meyer, Greissl \& Aversa}{Andersen
  et~al.}{2008}]{Andersen08}
Andersen M.,  Meyer M.~R.,  Greissl J.,    Aversa A.,  2008, ApJL, 683, L183

\bibitem[\protect\citeauthoryear{Bastian, Covey \& Meyer}{Bastian
  et~al.}{2010}]{Bastian10}
Bastian N.,  Covey K.~R.,    Meyer M.~R.,  2010, ARA\&A, 48, 339

\bibitem[\protect\citeauthoryear{Bate}{Bate}{2009}]{Bate09}
Bate M.~R.,  2009, MNRAS, 392, 590

\bibitem[\protect\citeauthoryear{Brice{\~n}o, Hartmann, Stauffer \&
  Martin}{Brice{\~n}o et~al.}{1998}]{Briceno98}
Brice{\~n}o C.,  Hartmann L.,  Stauffer J.,    Martin E.,  1998, AJ, 115, 2074

\bibitem[\protect\citeauthoryear{Brice{\~n}o, Luhman, Hartmann, Stauffer \&
  Kirkpatrick}{Brice{\~n}o et~al.}{2002}]{Briceno02}
Brice{\~n}o C.,  Luhman K.~L.,  Hartmann L.,  Stauffer J.~R.,    Kirkpatrick
  J.~D.,  2002, ApJ, 580, 317

\bibitem[\protect\citeauthoryear{Cartwright \& Whitworth}{Cartwright \&
  Whitworth}{2004}]{Cartwright04}
Cartwright A.,  Whitworth A.~P.,  2004, MNRAS, 348, 589

\bibitem[\protect\citeauthoryear{Dobashi, Uehara, Kandori, Sakurai, Kaiden,
  Umemoto \& Sato}{Dobashi et~al.}{2005}]{Dobashi05}
Dobashi K.,  Uehara H.,  Kandori R.,  Sakurai T.,  Kaiden M.,  Umemoto T.,
  Sato F.,  2005, PASJ, 57, 1

\bibitem[\protect\citeauthoryear{Gomez, Hartmann, Kenyon \& Hewitt}{Gomez
  et~al.}{1993}]{Gomez93}
Gomez M.,  Hartmann L.,  Kenyon S.~J.,    Hewitt R.,  1993, AJ, 105, 1927

\bibitem[\protect\citeauthoryear{Goodwin, Hubber, Moraux \& Whitworth}{Goodwin
  et~al.}{2005}]{Goodwin05c}
Goodwin S.~P.,  Hubber D.~A.,  Moraux E.,    Whitworth A.~P.,  2005,
  {Astronomische Nachrichten}, 326, 1040

\bibitem[\protect\citeauthoryear{Goodwin \& Whitworth}{Goodwin \&
  Whitworth}{2007}]{Goodwin07b}
Goodwin S.~P.,  Whitworth A.~P.,  2007, A\&A, 466, 943

\bibitem[\protect\citeauthoryear{Gouliermis, Keller, Kontizas, Kontizas \&
  {Bellas-Velidis}}{Gouliermis et~al.}{2004}]{Gouliermis04}
Gouliermis D.,  Keller S.~C.,  Kontizas M.,  Kontizas E.,    {Bellas-Velidis}
  I.,  2004, A\&A, 416, 137

\bibitem[\protect\citeauthoryear{G{\"u}del, Briggs, Arzner, Audard, Bouvier,
  Feigelson, Franciosini, Glauser, Grosso, Micela, Monin, Montmerle, Padgett,
  Palla, Pillitteri, Rebull, Scelsi, Silva, Skinner, Stelzer \&
  Telleschi}{G{\"u}del et~al.}{2007}]{Guedel07}
G{\"u}del M.,  Briggs K.~R.,  Arzner K.,  Audard M.,  Bouvier J.,  Feigelson
  E.~D.,  Franciosini E.,  Glauser A.,  Grosso N.,  Micela G.,  Monin J.,
  Montmerle T.,  Padgett D.~L.,  Palla F.,  Pillitteri I.,  Rebull L.,  Scelsi
  L.,  Silva B.,  Skinner S.~L.,  Stelzer B.,    Telleschi A.,  2007, A\&A,
  468, 353

\bibitem[\protect\citeauthoryear{Guieu, Dougados, Monin, Magnier \&
  Mart{\'i}n}{Guieu et~al.}{2006}]{Guieu06}
Guieu S.,  Dougados C.,  Monin J.-L.,  Magnier E.,    Mart{\'i}n E.~L.,  2006,
  A\&A, 446, 485

\bibitem[\protect\citeauthoryear{Gutermuth, Megeath, Myers, Allen \&
  Fazio}{Gutermuth et~al.}{2009}]{Gutermuth09}
Gutermuth R.~A.,  Megeath S.~T.,  Myers P.~C.,  Allen L.~E.,    Fazio J. L. P.
  G.~G.,  2009, ApJS, 184, 18

\bibitem[\protect\citeauthoryear{Joergens}{Joergens}{2006}]{Joergens06}
Joergens V.,  2006, A\&A, 448, 655

\bibitem[\protect\citeauthoryear{Kenyon, Dobrzycka \& Hartmann}{Kenyon
  et~al.}{1994}]{Kenyon94}
Kenyon S.~J.,  Dobrzycka D.,    Hartmann L.,  1994, AJ, 108, 1872

\bibitem[\protect\citeauthoryear{Kenyon, G{\'o}mez \& Whitney}{Kenyon
  et~al.}{2008}]{Kenyon08}
Kenyon S.~J.,  G{\'o}mez M.,    Whitney B.~A.,  2008, {Low Mass Star Formation
  in the Taurus-Auriga Clouds}.
pp 405--+

\bibitem[\protect\citeauthoryear{Kenyon \& Hartmann}{Kenyon \&
  Hartmann}{1995}]{Kenyon95}
Kenyon S.~J.,  Hartmann L.,  1995, ApJS, 101, 117

\bibitem[\protect\citeauthoryear{Kirk \& Myers}{Kirk \& Myers}{2010}]{Kirk10}
Kirk H.,  Myers P.~C.,  2010, ApJ, accepted (arXiv: 1011.1416)

\bibitem[\protect\citeauthoryear{Kirk, {Ward-Thompson} \& Andr{\'e}}{Kirk
  et~al.}{2005}]{Kirk05}
Kirk J.~M.,  {Ward-Thompson} D.,    Andr{\'e} P.,  2005, MNRAS, 360, 1506

\bibitem[\protect\citeauthoryear{Kroupa}{Kroupa}{2002}]{Kroupa02}
Kroupa P.,  2002, Science, 295, 82

\bibitem[\protect\citeauthoryear{Kroupa \& Bouvier}{Kroupa \&
  Bouvier}{2003}]{Kroupa03b}
Kroupa P.,  Bouvier J.,  2003, MNRAS, 346, 343

\bibitem[\protect\citeauthoryear{Littlefair, Naylor, Jeffries, Devey \&
  Vine}{Littlefair et~al.}{2003}]{Littlefair03}
Littlefair S.~P.,  Naylor T.,  Jeffries R.~D.,  Devey C.~R.,    Vine S.,  2003,
  MNRAS, 345, 1205

\bibitem[\protect\citeauthoryear{Luhman}{Luhman}{2000}]{Luhman00}
Luhman K.~L.,  2000, ApJ, 544, 1044

\bibitem[\protect\citeauthoryear{Luhman}{Luhman}{2004}]{Luhman04}
Luhman K.~L.,  2004, ApJ, 617, 1216

\bibitem[\protect\citeauthoryear{Luhman}{Luhman}{2006}]{Luhman06}
Luhman K.~L.,  2006, ApJ, 645, 676

\bibitem[\protect\citeauthoryear{Luhman, Allen, Espaillat, Hartmann \&
  Calvet}{Luhman et~al.}{2010}]{Luhman10}
Luhman K.~L.,  Allen P.~R.,  Espaillat C.,  Hartmann L.,    Calvet N.,  2010,
  ApJS, 186, 111

\bibitem[\protect\citeauthoryear{Luhman, Brice{\~n}o, Stauffer, Hartmann,
  {Barrado y Navascu{\'e}s} \& Caldwell}{Luhman et~al.}{2003}]{Luhman03}
Luhman K.~L.,  Brice{\~n}o C.,  Stauffer J.~R.,  Hartmann L.,  {Barrado y
  Navascu{\'e}s} D.,    Caldwell N.,  2003, ApJ, 590, 348

\bibitem[\protect\citeauthoryear{Luhman, Mamajek, Allen \& Cruz}{Luhman
  et~al.}{2009}]{Luhman09}
Luhman K.~L.,  Mamajek E.~E.,  Allen P.~R.,    Cruz K.~L.,  2009, ApJ, 703, 399

\bibitem[\protect\citeauthoryear{Monin, Guieu, Rebull, Goldsmith, Fukagawa,
  M\'enard, Padgett, Stappelfeld, McCabe, Carey, {Noriega-Crespo}, Brooke,
  Huard, Terebey, Hillenbrand \& G{\"u}del}{Monin et~al.}{2010}]{Monin10}
Monin J.-L.,  Guieu S.,  Rebull L.,  Goldsmith P.,  Fukagawa M.,  M\'enard F.,
  Padgett D.,  Stappelfeld K.,  McCabe C.,  Carey S.,  {Noriega-Crespo} A.,
  Brooke T.,  Huard T.,  Terebey S.,  Hillenbrand L.,    G{\"u}del M.,  2010,
  A\&A, accepted (arXiv: 1004.2541)

\bibitem[\protect\citeauthoryear{Padoan \& Nordlund}{Padoan \&
  Nordlund}{2002}]{Padoan02}
Padoan P.,  Nordlund {\AA}.,  2002, ApJ, 576, 870

\bibitem[\protect\citeauthoryear{Padoan \& Nordlund}{Padoan \&
  Nordlund}{2004}]{Padoan04}
Padoan P.,  Nordlund {\AA}.,  2004, ApJ, 617, 559

\bibitem[\protect\citeauthoryear{Palla \& Stahler}{Palla \&
  Stahler}{2002}]{Palla02}
Palla F.,  Stahler S.~W.,  2002, ApJ, 581, 1194

\bibitem[\protect\citeauthoryear{Parker \& Goodwin}{Parker \&
  Goodwin}{2007}]{Parker07}
Parker R.~J.,  Goodwin S.~P.,  2007, MNRAS, 380, 1271

\bibitem[\protect\citeauthoryear{Prim}{Prim}{1957}]{Prim57}
Prim R.~C.,  1957, Bell Syst. Tech. J., 36, 1389

\bibitem[\protect\citeauthoryear{Reipurth \& Clarke}{Reipurth \&
  Clarke}{2001}]{Reipurth01}
Reipurth B.,  Clarke C.~J.,  2001, AJ, 122, 432

\bibitem[\protect\citeauthoryear{Sabbi, Sirianni, Nota, Tosi, Gallagher, Smith,
  Angeretti, Meixner, Oey, Walterbos \& Pasquali}{Sabbi et~al.}{2008}]{Sabbi08}
Sabbi E.,  Sirianni M.,  Nota A.,  Tosi M.,  Gallagher J.,  Smith L.~J.,
  Angeretti L.,  Meixner M.,  Oey M.~S.,  Walterbos R.,    Pasquali A.,  2008,
  AJ, 135, 173

\bibitem[\protect\citeauthoryear{Sana, Momany, Gieles, Carraro, Beletsky,
  Ivanov, {De Silva} \& James}{Sana et~al.}{2010}]{Sana10}
Sana H.,  Momany Y.,  Gieles M.,  Carraro G.,  Beletsky Y.,  Ivanov V.~D.,  {De
  Silva} G.,    James G.,  2010, A\&A acccepted, (arXiv: 1003.2208)

\bibitem[\protect\citeauthoryear{{Schmalzl}, {Kainulainen}, {Quanz}, {Alves},
  {Goodman}, {Henning}, {Launhardt}, {Pineda} \&
  {Rom{\'a}n-Z{\'u}{\~n}iga}}{{Schmalzl} et~al.}{2010}]{Schmalzl10}
{Schmalzl} M.,  {Kainulainen} J.,  {Quanz} S.~P.,  {Alves} J.,  {Goodman}
  A.~A.,  {Henning} T.,  {Launhardt} R.,  {Pineda} J.~E.,
  {Rom{\'a}n-Z{\'u}{\~n}iga} C.~G.,  2010, ApJ, accepted (arXiv: 1010.2755)

\bibitem[\protect\citeauthoryear{Siess, Dufour \& Forestini}{Siess
  et~al.}{2000}]{Siess00}
Siess L.,  Dufour E.,    Forestini M.,  2000, A\&A, 358, 593

\bibitem[\protect\citeauthoryear{Stamatellos, Hubber \& Whitworth}{Stamatellos
  et~al.}{2007}]{Stamatellos07b}
Stamatellos D.,  Hubber D.~A.,    Whitworth A.~P.,  2007, MNRAS, 382, L30

\bibitem[\protect\citeauthoryear{Thies \& Kroupa}{Thies \&
  Kroupa}{2007}]{Thies07}
Thies I.,  Kroupa P.,  2007, ApJ, 671, 767

\bibitem[\protect\citeauthoryear{Ungerechts \& Thaddeus}{Ungerechts \&
  Thaddeus}{1987}]{Ungerechts87}
Ungerechts H.,  Thaddeus P.,  1987, ApJS, 63, 645

\bibitem[\protect\citeauthoryear{Walter \& Boyd}{Walter \&
  Boyd}{1991}]{Walter91}
Walter F.~M.,  Boyd W.~T.,  1991, ApJ, 370, 318

\bibitem[\protect\citeauthoryear{Whitworth, Bate, Nordlund, Reipurth \&
  Zinnecker}{Whitworth et~al.}{2007}]{Whitworth07}
Whitworth A.~P.,  Bate M.~R.,  Nordlund {\AA}.,  Reipurth B.,    Zinnecker H.,
  2007, in Reipurth B.,  Jewitt D.,   Keil K.,  eds, {Protostars and Planets V}
  {The Formation of Brown Dwarfs: Theory}.
pp 459--476

\bibitem[\protect\citeauthoryear{Whitworth, Stamatellos, Walch, Kaplan,
  Goodwin, Hubber \& Parker}{Whitworth et~al.}{2010}]{Whitworth10}
Whitworth A.~P.,  Stamatellos D.,  Walch S.,  Kaplan M.,  Goodwin S.,  Hubber
  D.,    Parker R.,  2010, in {R.~de Grijs \& J.~R.~D.~L{\'e}pine} ed., IAU
  Symposium Vol.~266 of IAU Symposium, {The formation of brown dwarfs}.
pp 264--271

\end{thebibliography}

\label{lastpage}

\end{document}